\documentclass[preprint2]{aastex}

\usepackage{amstex,epsfig}

\newcommand{\fb}{$f_\mathrm{B}$}

\def\eg{{\it e.g.,}\,}
\def\ie{{\it i.e.,}\,}

\def\whz{{W Hz$^{-1}$}}

\begin{document}

\title{Radio-selected Galaxies in Very Rich Clusters at $z \le$ 0.25: 
II.~Radio Properties and Analysis}

\author{Glenn \,E.\ Morrison}
\affil{California Institute of Technology, IPAC, M/S\,100-22,
Pasadena, CA 91125; morrison@@ipac.caltech.edu}
\author{ Frazer \,N.\ Owen}
\affil{National Radio Astronomy Observatory\altaffilmark{1}, P. O. Box O, Socorro,
NM 87801; fowen@@aoc.nrao.edu}

\altaffiltext{1}{The National Radio Astronomy Observatory is operated 
by Associated Universities, Inc., under a cooperative agreement with
the National Science Foundation.}

\begin{abstract}

We report on the properties of radio-selected galaxies within 30
very-rich Abell clusters with $z \la\,0.25$.  The radio, optical, and
x-ray data for these clusters were presented in Paper I
\citep{mor00}. These radio data sample the ultra-faint ($L_{1.4} 
\ge\,2\times 10^{22}$ W Hz$^{-1}$) radio galaxy population with
$M_{\mathrm{R}} \le$\,$-21$ using the well-known FIR/radio correlation
to link the radio with ongoing star formation within individual
cluster galaxies. Spectroscopic redshifts exist for $\sim$\,96$\%$ of
the optical identifications. These radio-selected galaxies reveal the
`active' galaxy population (starburst and active galactic nuclei) within these rich
cluster environments that can be identified regardless of their level
of dust obscuration. These new radio data provide the largest sample
to date of low-luminosity radio galaxies within rich cluster
environments allowing an unbiased search for dusty starbursting
galaxies. For all clusters in our sample, we are sensitive to star
formation rates (M $\ge$ 5M$_{\sun}$) $\ga$\,5\,M$_{\sun}$yr$^{-1}$.
We have found that the excess number of low-luminosity `starburst'
radio-selected galaxies (SBRG) found by \cite{owe99} in Abell 2125 is
{\it not} indicative of other rich clusters in our sample. The average
fraction of SBRG is $\langle\,f_{\mathrm{SBRG}}\,\rangle =
0.022\pm0.003$. The A2125 fraction is $f_{\mathrm{SBRG}} =
0.09\pm0.03$ which is significantly different from the sample average
at a $>$\,99.99$\%$ confidence level. Both A1278 and A1689 are
slightly different from the rest of the sample at $\sim$\,90$\%$
confidence level. The bimodal structure of both the x-ray brightness
distribution and optical adaptively smoothed images of A1278 and A2125
suggests that ongoing cluster-cluster mergers may be enhancing this
SBRG population. The A1689 excess low-luminosity (and high-luminosity)
radio galaxy population may be due to interaction with the ICM. The
mid-infrared ISOCAM results for A1689's radio galaxy population
suggests that the radio emission for both low- and high-luminosity
radio galaxies is AGN in origin except for one radio galaxy.  There is
a significant spatial distribution difference between the low and
high-luminosity (HLRG) radio-selected populations. The SBRG have a
core radius of $0.40\pm0.08$~Mpc which is $>$\,$3\times$ larger than
the HLRG core radius. In addition, 48$\%$ of the SBRGs have colors
that are bluer than a typical Sab galaxy compared to 4$\%$ for the
HLRGs. The average absolute magnitude for the SBRG's is
$\langle\,M_{\mathrm{R}}\,\rangle =-21.93\pm0.05$, while for the
HLRG's it is $\langle\,M_{\mathrm{R}}\,\rangle =-22.33\pm0.07$,
indicating that the SBRG are less optically luminous than their HLRG
counterparts. The HLRGs seem to be a subclass of the cluster's massive
red elliptical population, while the SBRGs have a projected radial
distribution more like the blue spiral population. Our results
indicate that most of the SBRGs are probably gas-rich disk galaxies
undergoing $\ga$\,5\,M$_{\sun}$\,yr$^{-1}$ of star-formation.

\end{abstract}


\section{Introduction}

Hierarchical clustering models predict that clusters of galaxies are
assembled by a continuous coalescence of subclusters
\citep[e.g.,][]{whi76,evr90b,lac93,lac94,bal98a}. The ongoing mass 
accretion yields artifacts in the form of cluster asymmetry and
significant amounts of subclustering. The remnants of structure
formation are visible in current epoch clusters as substructure and
cluster-cluster merging events. Estimates of detectable substructure
in rich nearby clusters range from 30$\%$-40$\%$
\citep[e.g.,][]{gel82, wes90}. The timescale for the accretion
events is $\sim$\,1\,Gyr for group-cluster encounters and
$\sim$\,4\,Gyr for cluster-cluster mergers\citep{evr90b}, compared to
$\lesssim$0.3~Gyr for a typical starburst (SB)\citep[e.g.,][]{ken98}.
Such SB episodes show up as a cluster-wide enhancement of the
radio-selected galaxy population. Therefore, while the substructure
caused by large-scale-structure  formation will be visible for a
significant fraction of a Hubble time, the cluster-wide enhanced radio
activity, as in the case of A2125
\citep{owe99}, will be rather brief.

The Butcher-Oemler (BO) effect \citep{but84b} has been suggested as a
link between galaxy evolution and cluster dynamics. The hierarchical
models predict that distant clusters assemble from smaller subclumps
at a much higher rate than similar mass nearby clusters. These smaller
groups have a higher fraction of gas-rich, late-type galaxies, thereby
providing a distant cluster with a population that would support
significant amounts of induced star formation (SF). In fact, these
late-type galaxies may be transformed into early-type galaxies by
environmental dynamics of the cluster \citep[e.g.,][]{oem97}.

The \cite{cal97} spectral study of early-type galaxies in rich
clusters with significant substructure shows that $\sim$\,15$\%$ of
these galaxies have signatures of current or ongoing SF. This suggests
that cluster-group merger activity signified by the substructure may
alter the star formation rate (SFR) of a cluster galaxy
population. Induced SF may result from shocks caused by the collision
between the intracluster medium of the cluster and that of the
group. Alternatively, the main driver of this activity may be
galaxy-galaxy collisions in regions where the velocity dispersion is
low enough and the density is high enough for such encounters to be
efficient. The different regions that the starburst population
inhabits within the rich cluster gives us clues as to the triggering
mechanism for this activity. Radio observations allow us to identify
this population without being hampered by dust extinction or the
K-correction.

This paper is the second in a series studying the ultra-faint
radio-selected galaxy populations associated with rich clusters of
galaxies as a function of redshift.  We analyze both the high and low
luminosity radio-selected galaxy population within 30 very rich Abell
clusters (richness $\ge$2) with $z \la 0.25$. The data used in our
analysis was presented in paper I \citep{mor00}.

Initially, \cite{dwa99} conducted a VLA 20 cm wide-field continuum
study of two very rich Abell clusters, A2125 and A2645, looking at the
radio galaxy population down to $\sim$2$\times$10$^{22}$ W
Hz$^{-1}$. The reason for choosing this lower radio luminosity limit
is that \cite{con91} had determined that in the local universe this
limit has a statistically higher fraction of starburst-powered
radio-selected galaxies and few galaxies whose radio emission is
powered by AGNs. Both clusters are at the same redshift (z=0.25) and
richness (R=4), but A2125 has a higher blue galaxy population
($f_{\mathrm{B}}$ = 0.19) than A2645 ($f_{\mathrm{B}}$ = 0.03). The
radio galaxy population is also substantially different: A2125 has 26
radio detected galaxies whereas A2645 has 4 \citep{owe99}.

Most of the radio galaxy population of A2125 is not confined to the
core region of the cluster ($r <$ 400 kpc) but is distributed out to a
radius of 2.5 Mpc. The radio-selected galaxies in the cluster core are
red in color, with none of the blue Butcher-Oemler galaxies detected
down to the radio luminosity limit. The origin for the radio emission
is both AGN and massive ongoing SF. The latter occurs more frequently
in the outer cluster region of A2125.

This result raises several interesting questions. Are we just seeing
the infall of the field population at $z = 0.25$, or are we witnessing
an enhancement in the activity level of the population due to the
cluster environment?  Do all very rich clusters show enhanced activity
in the form of starburst and AGNs, or only the ones at higher
redshift?

To understand if this is a richness or a redshift effect, we have
constructed a sample of 34 very rich clusters from 0.02 $\la z \la$
0.41 to study how the radio properties of these clusters evolve (if at
all) over the last $\sim$5 Gyr.  VLA radio observations at 20\,cm
offer a wide field-of-view ($\sim$30 arcmin), allowing us to sample
out to 2.5 Mpc radius from the cluster core. Nearby clusters ($z \leq
0.06$) were analysed using the NRAO VLA Sky Survey (NVSS) which
provided the required linear coverage (5\,Mpc diameter search area)
and sensitivity.  At this projected distance from the cluster core we
should detect any infalling starbursting field galaxies. Our detection
limit of 2$\times$10$^{22}$ W Hz$^{-1}$ ($H_0$ =
75\,km\,s$^{-1}$\,Mpc$^{-1}$; $q_0 = 0.1$) yields a SFR(M $\ge$
5M$_{\sun}$) sensitivity limit of $\ga 5$ M$_{\sun}$ yr$^{-1}$
assuming the SFR relation from
\cite{con92}.

In this paper, we analyze the radio properties of 30 of the richest,
nearest Abell clusters from 0.02 $\la$ z $\la$ 0.25. These clusters
were chosen to be as rich or richer than the four $z = 0.4$ clusters.
We investigate the characteristics of the different radio galaxy
populations and their relationship to the cluster properties, such as
richness, core richness, compactness, and x-ray luminosity. In
addition, the Butcher-Oemler-defined blue color and absolute optical
luminosity of the radio-selected objects are also explored.

\section{The Radio-selected Galaxy Classes}

In this study we were unable to spectroscopically classify
radio-selected galaxies as starbursts or AGNs given the quality of the
spectra. However, useful redshift measurements were obtained. The new
spectroscopic data and resulting spectroscopic classifications of the
radio-selected galaxies will be discussed in paper IV
\citep{mor00c}. Thus, we relied on statistical means to separate the
galaxies into these two classes.  Based on Figure 9 from
\citeauthor{con89}'s \citeyearpar{con89} paper showing the 
local field galaxy radio luminosity function (RLF), we see that a
luminosity of $10^{23}$~W Hz$^{-1}$ divides the starburst-spiral
population from the AGN-E/S0 population. Radio-selected galaxies with
luminosities below this limit are statistically more likely to have
their radio emission powered by non-thermal synchrotron emission as a
by-product of massive SF. Table~\ref{table_1} defines the
different radio luminosity classes. The different radio galaxy classes
are high-luminosity (HLRG), low-luminosity (LLRG), and `starburst'
(SBRG). Most of the analysis will be concerned with the HLRG and the
SBRG classes.

The SBRG class was created from the LLRG class because of the excess
of early-type galaxies within galaxy clusters \citep{oem74}. As we can
see in Condon's RLF plot, while the elliptical/AGN population drops
below the starburst-spiral population at $10^{23}$~W Hz$^{-1}$, AGN
powered radio galaxies below this break still exist. Given the larger
population of early-type galaxies that inhabit clusters, we have
chosen a lower break in the RLF at $10^{22.75}$~W Hz$^{-1}$. Based on
the RLF, our SBRG should have a lower level of AGN contamination than
the LLRG class. Therefore, statistically the SBRGs should be primarily
powered by starbursts and the HLRGs by AGNs.


\section{Spatial Distribution of Radio-selected Galaxies}

The radial distribution with respect to the cluster center of the
high- and low-luminosity radio galaxies allows us to examine how the
different radio galaxy populations are distributed with respect to
each other and the cluster environment.  This may give us clues to
their past history.  Do low-luminosity radio-selected galaxies belong
to some sort of subclass of the cluster population, \eg recently
infalling star forming blue field galaxies or are they cluster blue
galaxies?

\subsection{Luminosity Class Spatial Distribution}\label{SPATIAL_DISTRIBUTION}

We begin our examination with the spatial radial distribution of the
radio-selected galaxies within the cluster (verified by spectroscopic
redshift) as a function of projected distance from the cluster center.
Our radio sample is complete out to a linear projected distance of 2.5
Mpc.  Figure~\ref{FIG_1} shows the radio galaxy surface-density
plotted against the projected distance from the center of the
cluster. Thirty clusters were used. The centers are based on the X-ray
center fits, with errors $\pm$ 15-25\arcsec \, \citep{mor00}. The
surface density, $\sigma\,(\#/Mpc^{2})$, was calculated by dividing
the number of detections occurring in each bin by the annular search
area in Mpc$^2$. The area of each annulus was multiplied by the number
of surveyed clusters used to make the sample.

\placefigure{fig_1}

The distribution of the radio galaxies is rather striking. The high
luminosity radio galaxies (HLRG) are very clustered near the cluster
cores. This is consistent with the \cite{led95b,led95a} survey of 293
clusters, which showed that HLRGs (mostly FR I's) preferentially
reside in the centers of rich clusters because the most optically
luminous galaxies occur there.

The distribution of the SBRGs is a {\it new} result. Radio galaxies
with luminosities between $10^{22.3-22.75}$ W Hz$^{-1}$ seem to avoid
the cluster centers and distribute themselves more widely (for $r <
1.0$\,Mpc), similar to blue galaxies in clusters. A Kolmogorov-Smirnov
test between the spatial distributions of HLRGs and SBRGs indicates
that they are from two separate parent populations at the $> 99\%$
confidence level.  The SBRG's follow the more widely distributed
cluster spiral population, while the HLRG's appear to trace the
massive red E/S0 cluster population.

\subsection{King Model Comparison}

Clusters that are highly symmetric in projected shape and have a high
concentration at the center or core are typically called ``regular''
and, in some cases, are believed to be virialized. These clusters are
well-fit by the King model, represented by
equation~\ref{KING_DISTRIBUTION}. This model is an analytical solution
to the inner part of an isothermal function \citep{kin62}:

\begin{equation}
\Sigma(r) = \frac{\Sigma_0}{1 + (r/r_{\mathrm{c}})^2}.
\label{KING_DISTRIBUTION}
\end{equation}
\vspace{2 mm}

\placefigure{fig_2}

Figure~\ref{FIG_2} shows an overabundance of high-luminosity radio
galaxies near the cluster core, while figure~\ref{FIG_3} shows that
the low-luminosity radio galaxies are more widely distributed than
predicted, based on a virialized cluster. All radio-selected galaxies
used in these plots have measured redshifts that are consistent with
their cluster redshift. The dispersion or width measure seen in these
plots is roughly defined by the core radius, $r_c$. The core radius is
where the projected surface density is half the central density,
$\Sigma_0$. The core radius is a function of cluster morphology
\citep{sar85}, but several studies \citep{bah75, gir95, ada98} have found
values of $r_{\mathrm{c}}$ that are consistent with each other,
yielding a value of $\sim$0.2 Mpc for regular clusters.

\placefigure{fig_3}

Assuming a King distribution for the radio galaxies, we determined the
values of the core radii, $r_{\mathrm{c}}$, that are compatible with
the SBRGs and the HLRGs distributions.  The integrated or cumulative
King function,

\begin{equation}
\sigma(r) = \pi r_c^2 \ln(1 + (r/r_c)^2),
\label{CUMULATIVE_KING_FNC}
\end{equation}
\vspace{2 mm}

\noindent which was normalized to one at r = 2.5 Mpc, was used with the
K-S test to determine the $r_c$ values for low- and high-luminosity
radio-selected galaxy distribution. The results show that the SBRG
population does not come from a King distribution with a small core
($r_c \sim 0.15$\,Mpc) radius that would fit the luminous radio
population at a confidence level $>$\,99$\%$.

The $r_{\mathrm{c}}$ derived values from a $\chi^2$ fit to the King
function for the different radio populations are $r_{\mathrm{c}}=
0.40\pm0.08$~Mpc and $r_{\mathrm{c}}= 0.12\pm0.02$~Mpc for the SBRGs
and HLRGs, respectively.  \cite{mor00b} has found that
$r_{\mathrm{c}}=0.26\pm0.11$\,Mpc for the red population of regular
clusters, which is similar to the $r_{\mathrm{c}}$ value for the
HLRGs, suggesting that they are a subclass of a cluster's massive red
population. \citeauthor{led95a}'s \citeyearpar{led95a} much larger
sample shows this result much more strongly for the HLRGs.
\cite{mor00b} also found that the blue population for the compact
clusters have $r_{\mathrm{c}}= 0.6\pm0.23$~Mpc which is within the
errors of the $r_{\mathrm{c}}$ value for SBRGs. Based on the SBRG
radio luminosity and core radius value these galaxies seem to be a
subclass of the cluster's blue galaxy population. Further evidence of
this is given in section~\ref{CRSG} and in figure~\ref{FIG_5}ZZZ.

\section{Radio Galaxy Fractions}

The radio galaxy fraction of a cluster was defined in Paper
I. Briefly, the fraction of radio galaxies, $f_{\mathrm{RG}}$, is the
number of radio galaxies normalized by the total number of galaxies,
$N_{2.5}$, sampled (corrected for the background). $N_{2.5}$ is the
number of galaxies within 2.5\,Mpc of the cluster center with an
absolute $R$ magnitude cutoff of $-21$. The radio galaxy fraction for
each cluster with $n$ total galaxies is a binomial probability. We
determined how this probability differs in a given cluster from a
``true'' probability derived from the sample as a whole. This imposes
a constraint on the system reducing the degrees of freedom by
one. From this method, we constructed 1$\sigma$ errors for the
$f_{\mathrm{RG}}$ values. Significant levels for clusters which have
radio galaxy fractions inconsistent with the rest of the sample were
calculated using the same method.

Table~\ref{table_1} shows the radio-selected galaxy classes we will
study. Given the higher contamination of AGNs in the LLRG class we
will analyze only the SBRG and HLRG populations. These two
radio-selected populations will be investigated separately, in order
to tell if any clusters show an excess fraction relative to the other
rich clusters. If a large enhancement exists for a particular cluster,
this will provide evidence that ongoing physical processes within the
cluster environment may be causing these galaxies to have increased
radio emission.

Given the uncertainty in the number of galaxies, $N_{2.5}$, within
2.5\,Mpc of the cluster core and having $M_{\mathrm{R}} \leq -21$ for
clusters with $z < 0.1$ \citep[see][]{mor00b}, we will restrict our
radio fraction analysis to clusters with $z > 0.1$.

\placefigure{fig_4}

\subsection{The Radio-selected  Population}

We are now in a position to answer the following question: Is the
radio galaxy population of A2125 ubiquitous for all rich clusters?
Figure~\ref{FIG_4} shows the SBRG $f_{\mathrm{RG}}$ as a solid line
for all $0.1 \la z\la 0.25$ clusters. A2125 has the largest fraction
of SBRGs\footnote{That is this cluster deviates from the rest of the
sample by having an excess of galaxies in this range of radio
luminosity.}; $f_{\mathrm{SBRG}}~= 0.09\pm0.02$ which is significantly
different from the rest of the sample at the $>99.99\%$ confidence
level.  The mean for the SBRGs, $\langle\,f_{\mathrm{SBRG}}\,\rangle$,
is 0.022$\pm$0.003. Other clusters: A1278 and A1689 (both have
$f_{\mathrm{SBRG}}~= 0.05\pm0.03$) are different from the mean only at
the $\sim\,90\%$ confidence level.  However, the large
$f_{\mathrm{RG}}$-value for A2125 is not a richness effect, since the
total number of radio-selected objects has been normalized by the
total number surveyed.

The spectroscopic classification of the radio-selected galaxies in
A2125 is described in \cite{owe99}. They find four classes of cluster
radio galaxies (old stellar population (OSP), starbursts, AGN, and
``intermediate'') defined by their optical spectra\footnote{In the
\cite{owe99} sample, characterization of the galaxies was done using
the emission and absorption lines of the spectra.} and colors.  50$\%$
(13/26) of these galaxies have spectral energy distributions (SED)
similar to OSP, based on their colors and/or 4000\AA \ break (D4000)
and lack of emission lines. These 13 OSP galaxies have their radio
luminosities divided into the following radio luminosity classes.
SBRGs make up 31$\%$, LLRGs 38$\%$, and HLRGs 62$\%$. All of the
radio-selected galaxies classified as starbursts in the \cite{owe99}
sample are SBRGs. The radio properties of their intermediate class,
defined by their bluer colors and/or small D4000, but weak or
undetected line emission, contain 88$\%$ SBRGs, 100$\%$ LLRGs, and no
HLRGs. This suggests that these objects may have a young stellar
component plus ongoing SF whose optical signature is hidden by
dust. The only spectroscopically-confirmed AGN is an HLRG. Thus,
75$\%$ of the SBRGs appear to have active SF (62$\%$ for the LLRGs),
while all the HLRGs appear to be AGN in nature.

The fraction of HLRGs in A2125 ($f_{\mathrm{RG}} = 0.03\pm0.01$) does
not display any excess with respect to the other rich clusters in the
sample as seen in Figure~\ref{FIG_4} where the HLRG radio galaxy
fractions are represented by the hatched pattern. However, A1689 shows
an enhanced $f_{\mathrm{RG}}$~value ($0.10\pm0.04$) compared to the
rest of the sample, which is significant at the $>$\,$99.9\%$
level. A1940 has a slightly higher $f_{\mathrm{HLRG}}$ value than the
rest of the sample at $0.06\pm0.03$ with a confidence level of $94\%$.
The mean for the HLRGs is $\langle\,f_{\mathrm{RG}}\,\rangle$ =
0.024$\pm$0.003.

\subsubsection{Colors of Radio-selected Galaxies}\label{CRSG}

In this section, we determine if a galaxy's optical color is dependent
on its radio luminosity and also provide qualitative evidence for the
probable power sources (AGN or starburst) for the radio emission.

After correcting for the color-magnitude (C-M) effect ({\em
e.g.} \,\cite{sta98}) and applying the K- correction to the
radio-selected galaxy's color, we compare the $(B-R)_{\mathrm{RG}}$
color of the galaxy to that of ``blue galaxies'' as defined by
\cite{but84b}. BO defines blue galaxies as having rest-frame $B-V$
colors at least 0.2 magnitudes bluer ($\Delta\,(B-V) =
0.2$)\footnote{Basically, the difference between the average $B-V$
color of a Sab galaxy and the average color of the red cluster
sequence.} than the E/S0 galaxy population.  Details can be found in
\cite{mor00b}.  In brief, in the rest-frame for a particular galaxy,
we calculated the average E/S0 color transformed to $B-R$, $\langle
B-R\rangle_{\mathrm{(E/S0)}}$, and the BO blue criterion,
$\Delta\,(B-R)$. Radio-selected galaxies that obey

\begin{equation}
\langle B-R\rangle_{\mathrm{(E/S0)}} - \Delta (B-R) \ge  (B-R)_{\mathrm{RG}}
\label{COLOR_SEP}
\end{equation}

\noindent are defined as blue,  where  $(B-R)_{\mathrm{RG}}$ is
the color index of the radio-selected galaxy. This criterion is the
\cite{but84b} definition of a blue galaxy transformed from the
$B-V$ color index to $B-R$.

The result for the SBRGs and the HLRGs is seen in
figure~\ref{FIG_5}. This plot separates at zero the red population
($<$ 0) and the blue population ($\ge$ 0). The color separation
indicates that most of the luminous radio galaxies are probably red
ellipticals, consistent with the well-known result that the host
galaxies of luminous radio sources are ellipticals and powered by AGNs
\citep[\eg][]{led95a}. If we selected galaxies with a radio luminosity
limit of $\leq 10^{23}$\,W Hz$^{-1}$, we find that 39$\%$ (17/44) of
these objects are blue, while the HLRG is only 4$\%$ (1/28)
blue. However, if we restrict the upper radio luminosity of the
galaxies to $10^{22.75}$\,W Hz$^{-1}$, \ie~the SBRG population,
thereby statistically selecting mostly starburst galaxies, this
results in 48$\%$ (15/31) of the galaxy population having colors that
are bluer than Sab galaxies. Thus, by restricting the upper radio
luminosity of the LLRGs to $10^{22.75}$\,W Hz$^{-1}$ we are thereby
statistically selecting mostly starbursting galaxies, where the radio
is powered by the ongoing SF. In the HLRG case, AGNs appear to be the
power source for the radio emission. In Paper IV, we will
spectroscopically classify the radio-selected objects thereby
determining their dominant power source for the radio emission.

\placefigure{fig_5}

\subsubsection{Absolute Magnitudes of Radio-selected Galaxies}

Figure~\ref{FIG_6} shows the absolute magnitude distribution of the
SBRG and the HLRG population.  The SBRGs have an
$\langle$$M_{\mathrm{R}}$$\rangle$ = $-21.93\,\pm$\,0.05 while the
HLRG have an $\langle$$M_{\mathrm{R}}$$\rangle$ = $-22.33\,\pm$\,0.07
indicating that the SBRGs are less optically luminous than the HLRGs.
Wilcoxon Rank test and the K-S test yield confidence levels $>99.9\%$
indicating that the SBRG and HLRG are from significantly different
populations, based on the absolute magnitude distribution.  These data
suggest that the SBRG are probably spirals, while the high-luminosity
radio sources are more likely to be massive cluster ellipticals.

\placefigure{fig_6}

\subsection{Cluster Environment}

What do these elevated $f_{\mathrm{SBRG}}$ values compared with the
rest of the sample, tell us?  Are they indicative of cluster
environmental effects on galaxy evolution? If cluster environment is
the cause of the radio enhancement, why do not all of our clusters
have similar $f_{\mathrm{SBRG}}$ values? Since the SBRGs are more
likely to be starburst galaxies, few rich cluster environments appear
to stimulate massive SF in their galaxies. Since `average' local
galaxies generally do not experience radio emission at $\ge
10^{22.3}$\,W Hz$^{-1}$ without having vigorous amounts of SF (SFR
$\ga 5$\,M$_{\sun}$\,yr$^{-1}$) or possessing an AGN
\citep{con92}, it may be that the  environment in these
few clusters is different and is somehow causing this radio weak
galaxy population.

Both A1278 and A2125 have bi-modal X-ray and adaptively-smoothed
optical number density distributions \citep[see][]{mor00}, which is
evidence for an ongoing cluster-cluster merger. Recent theoretical
work by \cite{bek99} suggests that the rapidly varying gravitation
potential of a group-cluster merger triggers a starburst in gas-rich
galaxies. The triggering is done by exciting the non-axisymmetric
structure of the galaxy, thereby funneling gas to the central region,
commencing a starburst. Thus, cluster mergers may stimulate radio
emission within gas-rich cluster members. However, other known
group-cluster mergers in this sample, such as A168, A754, A2111, and
A2256, do not show the same activity level in the radio
($\L_{1.4\mathrm{GHz}} \geq 10^{22.3}$ W Hz$^{-1}$) as is seen in
A1278 and A2125. \cite{tom96} looked for an enhanced blue fraction
(\fb) in A168 but found none.  One of their conclusions from this was
that the cluster members are gas deficient, thus unable to support
enhanced SFRs.

Another possible mechanism for the enhanced fraction of radio emitting
galaxies in clusters is galaxy-galaxy interactions and/or mergers.  In
A2125, \cite{led99} found that nearly 90$\%$ of both the red and blue
radio-selected galaxies appeared in pairs ($< 30$\,kpc projected
separation), compared with a pair fraction of $\sim$40$\%$ for other
cluster galaxies. Spectroscopically-measured relative velocities
indicate that only two of the 26 radio-selected galaxies (both
starbursts) have $\Delta\,V \la 300$\,km\,s$^{-1}$. In those two
cases, galaxy interactions are the likely trigger for the
starburst. The other pairs are comprised of chance superposition of
stars and galaxy pairs with $\Delta\,V \ge 500$\,km\,s$^{-1}$. These
high-speed galaxy encounters may support the galaxy harassment
scenario \citep{moo96}.

As for A1689, it is a relaxed cluster whose fraction of SBRGs could
not be enhanced by the cluster merger mechanism, but possibly enhanced
by pressure confinement from its substantial intracluster medium
(ICM). This also might be true for the high fraction of HLRGs found in
A1689. 

From \cite{duc02} mid-infrared ISOCAM\footnote{ISOCAM camera
\citep{ces96} onboard the ISO satellite.} study of A1689 cluster core
region, we find that out of the eleven radio-selected galaxies for
this cluster, three are outside the ISOCAM field-of-view (fov). Of the
remaining eight, three are not detected at 6.75$\mu$m ($LW$2 filter)
and 15$\mu$m ($LW$3 filter), two are detected at 6.75$\mu$m but not at
15$\mu$m, while the final three are detected in both bands. Following
the MIR classification criterion of \cite{duc02} we classify the radio
galaxies as follows.  For the three SBRGs in A1689 within ISOCAM's
fov, one is not detected in either band, one is classified as a
starburst, and the other as an AGN. For the five HLRGs within the
ISOCAM's fov, all are classified as AGNs. However, the majority of the
MIR galaxies detected by \cite{duc02} have $L_{\mathrm{1.4\,GHz}} <
2\times 10^{22}$ W\,Hz$^{-1}$ indicating that any hidden star
formation present in A1689 is below 5\,M$_{\sun}$ yr$^{-1}$ (M $\ge$
5M$_{\sun}$).

\subsubsection{Cluster Parameters}

\subsubsubsection{Cluster Richness and Compactness}

One important question is whether the richness (galaxy counts)
$N_{\mathrm{2.0}}$\footnote{$N_{\mathrm{2.0}}$ represents the number
of galaxies brighter than $M_{\mathrm{R}} = -20.5$ (roughly the number
between m$_{3}$ and m$_{3} + 2$.) within one Abell radius or
2.0\,Mpc. See \cite{mor00b} for details.} of a cluster correlates with
the fraction of radio galaxies. We did find a weak anti-correlation
between $f_{\mathrm{RG}}$~and $N_{2.0}$~at the 90$\%$ confidence
level, possibly suggesting that the detection rate for all radio
galaxies does not scale with the number of galaxies surveyed.
However, we found no significant ($\geq 2\sigma = 95\%$ confidence)
correlation of $f_{\mathrm{RG}}$, $f_{\mathrm{SBRG}}$, or
$f_{\mathrm{SBRG}}$ with richness ($N_{0.5}$ or $N_{\mathrm{2.0}}$).

Another question is whether the compactness of a cluster is correlated
to the radio galaxy fraction.  \cite{owe99} discuss the radio
population in A2125 and A2645, noting the excess of the SBRG
population in irregular cluster A2125, with the lack of such a
population in compact cluster A2645.  Their result suggests that the
large (21) SBRG population in A2125 may be driven by the ongoing
cluster-cluster merger or the coalescence of multiple subunits
\citep{wan97c}. A2645 has the appearance of a relaxed,
centrally-condensed compact cluster, whose population is dominated by
red galaxies. Given that we have a mixture of cluster types (\eg
~regular and irregular) in our sample, we will try to decouple the
population radio-selected galaxies from the cluster morphology by
using the following compactness parameter.

Compactness parameter, defined as
$\mathcal{C}$ = $N_{0.5}$/$N_{\mathrm{2.0}}$, is the ratio of the
\cite{bah81} counts, $N_{0.5}$, to the \cite{abe89}, galaxy counts,
$N_{\mathrm{2.0}}$. This ratio provides a rough quantitative measure
of a cluster's compactness or morphology (\ie ~regular-compact or
irregular-open). We found no significant ($\geq 2\sigma = 95\%$
confidence) correlation for the whole sample between
$f_{\mathrm{RG}}$, $f_{\mathrm{SBRG}}$, or $f_{\mathrm{SBRG}}$ with
cluster compactness. This negative result might be due to projection
effect on the sky where only well separated merging systems would have
a large $\mathcal{C}$. Or possibly the rather brief
($\lesssim$0.3~Gyr) period that cluster-wide starburst phase could
occur in such a cluster might have a large $\mathcal{C}$ but no large
fraction of SBRG.

\subsubsubsection{Cluster Blue Fraction}

\cite{mor00b} discusses our procedure for measuring
$f_{\mathrm{B}}$. In brief, $f_{\mathrm{B}}$\, is based on the
\cite{but84b}  definition, with the exception that we use a fixed
metric aperture of radius 0.5\,Mpc centered on the X-ray peak of the
cluster. The parameter $f_{\mathrm{B}}$ measures the blue galaxy
population in the cluster core with respect to the cluster's red
population, using a Sab galaxy as a fiducial point for defining the
blue population. The K-correction and the color-magnitude effect have
been applied to each cluster.

The $f_{\mathrm{B}}$ and $f_{RG}$ values of the clusters are not
correlated. This result is expected, given the different regions
sampled. The parameter $f_{\mathrm{B}}$ is measured over the central
core region of the cluster, whereas $f_{RG}$ is measured over a much
larger region, $r\le$ 2.5\,Mpc, as discussed in the last section. In
addition, the $f_{\mathrm{B}}$ magnitude limit is $R = -19$ compared
to $R = -21$ for the radio-selected galaxies. Moreover, given the SFR
threshold $\ga 5$\,M$_{\sun}$\,yr$^{-1}$ that can be detected in the
radio, typical late-type spiral galaxies
(SFR$\,\la\,$4\,M$_{\sun}$\,yr$^{-1}$) that would be detected by the
BO method would not be selected in the radio. However, the radio does
detect dust-enshrouded starbursting galaxies that would not be
selected by the BO method because of their rather red color.

\subsubsubsection{X-ray Luminosity}

The $f_{\mathrm{RG}}$ and $f_{\mathrm{SBRG}}$ values for all the
clusters show no significant correlation with the X-ray luminosities
of the clusters. Of interest is A2125 and A1278's $f_{\mathrm{SBRG}}$
values which have higher $f_{\mathrm{SBRG}}$ values at a similar X-ray
luminosity. However, too few clusters exist with bimodal X-ray/optical
distributions to draw any conclusions.

The high radio luminosity radio fraction ($f_{\mathrm{HLRG}}$) also
fails to demonstrate any significant correlation with the
$L_{\mathrm{x}}$ of the cluster. There is an indication that a
pressure confinement enhancement effect may be taking place in
A1689. This cluster has a high X-ray luminosity, as well as the
highest $f_{\mathrm{HLRG}}$ fraction of any of the $z \la 0.25$
clusters.  While this is only one cluster, it does support the idea
that pressure from the intracluster material could increase the radio
luminosity of twin jets near the cluster center, where these HLRG
sources are located.

\section{Conclusion}

In this paper we have studied the radio-selected galaxy population of
a subsample of very rich ($R \ge 2$) Abell galaxy clusters with $z \le
0.25$.  The weak anti-correlation between $f_{\mathrm{RG}}$~and
$N_{2.0}$~at the 90$\%$ confidence level suggests possibly that the
detection rate for all radio galaxies does not scale with the number
of galaxies surveyed.  However, overall, we found no significant
($\geq 2\sigma = 95\%$ confidence) correlation or anti-correlation
of $f_{\mathrm{RG}}$, $f_{\mathrm{SBRG}}$, or $f_{\mathrm{SBRG}}$ with
cluster richness ($N_{2.0}$ or $N_{\mathrm{0.5}}$), compactness, blue
fraction, or x-ray.

There are only a few clusters that have radio galaxy fractions (SBRG
and HLRG) that are inconsistent with the rest of the sample.  We find
the following average radio galaxy fractions:
$\langle\,f_{\mathrm{SBRG}}\,\rangle = 0.022\pm0.003$ and
$\langle\,f_{\mathrm{HLRG}}\,\rangle = 0.024\pm0.003$.  The cluster
with an excess at the $>~99.99\%$ confidence level of SBRGs is A2125,
with a $f_{\mathrm{SBRG}} = 0.09\pm0.03$. This is {\it not} a
richness-induced effect as we normalized by the number of galaxies
sampled.  Two clusters with weak $\sim 90\%$ confidence level
deviations from $\langle\,f_{\mathrm{SBRG}}\,\rangle$ are A1278 at
$0.05\pm0.03$ and A1689, at $0.05\pm0.03$.

The bimodal structure in the optical and X-ray for A1278 and A2125
suggests that a possible cluster-cluster merger may be driving this
excess in SBRGs. The mechanism for the increased fraction of SBRGs in
A1689 may be the result of the ISM of the galaxies being compressed by
their passage through the ICM. All the SBRGs in A1689 are within a
projected distance of 0.8\,Mpc from cluster core.

The large fraction of HLRGs in A1689 ($0.09\pm0.04$) and A1940
($0.06\pm0.03$) may also be a result of the cluster environment. The
HLRGs close projected distance from the cluster center ($< 0.8$Mpc for
both clusters) suggests that this population may be enhanced due to
pressure confinement by the ICM. The MIR ISOCAM results for A1689's
radio galaxy population suggest most are AGN in nature lends support
to this idea.

The different spatial distribution between the SBRGs and the HLRGs is
one of the most significant results of this paper. The core radius
values of the SBRGs is $r_{\mathrm{c}} = 0.40\pm0.08$~Mpc. These are a
factor $>$\,$3\times$ larger than the average value of
$r_{\mathrm{c}}= 0.12\pm0.02$~Mpc for HLRGs. The large difference in
the fitted core radii values of the SBRGs and the HLRGs indicates a
strong difference exists between their representative populations. The
HLRGs are probably a subclass of the cluster's massive {\it red}
elliptical population, while the SBRGs have a distribution more like
the {\it blue} spiral population.
 
The difference in $\langle\,M_{\mathrm{R}}\,\rangle$ values between
the SBRGs and the HLRGs indicates an absolute magnitude segregation
between the two populations, with the higher optical luminosity
galaxies belonging to the HLRGs.  This is in agreement with the core
radius results for the SBRGs and HLRGs. Also the colors of the three
radio populations suggest that the SBRGs have colors that are much
bluer than the HLRGs.  These results indicate that a large fraction of
the SBRGs are probably gas-rich disk galaxies with SFR
$\ga$5\,M$_{\sun}$\,yr$^{-1}$. It is unclear as to what triggered the
SF in the SBRGs for most clusters. It must be noted that contamination
due to AGNs in the SBRGs has not been completely removed. Paper IV
will cover the spectroscopic classifications of the radio-selected
galaxies.

A larger, more morphologically diverse sample is currently being
studied that contains more irregular rich clusters similar to A1278
and A2125. This will allow us to decouple the effects that cluster
dynamics have on the radio properties of cluster galaxies.

\acknowledgments

We thank an anonymous referee for helpful comments.  We would also
like to thank Carol Lonsdale, Neal Miller, Michael Ledlow, Bill Keel,
and Schuyler Van Dyk.  G.E.M. acknowledges financial support from the
NRAO predoctoral fellowship program during which most of this work was
completed. G.E.M. also gratefully acknowledges the support of NASA-JPL
(contract $\#$1147-001 and $\#$1166) and also the support of Vanguard
Research, Inc.



\begin{thebibliography}{38}
\expandafter\ifx\csname natexlab\endcsname\relax\def\natexlab#1{#1}\fi

\bibitem[{{Abell} {et~al.}(1989){Abell}, {Corwin}, \& {Olowin}}]{abe89}
{Abell}, G.~O., {Corwin}, H.~G., J., \& {Olowin}, R.~P. 1989, \apjs, 70, 1

\bibitem[{{Adami} {et~al.}(1998){Adami}, {Mazure}, {Katgert}, \&
  {Biviano}}]{ada98}
{Adami}, C., {Mazure}, A., {Katgert}, P., \& {Biviano}, A. 1998, \aap, 336, 63

\bibitem[{{Bahcall}(1975)}]{bah75}
{Bahcall}, N.~A. 1975, \apj, 198, 249

\bibitem[{{Bahcall}(1981)}]{bah81}
---. 1981, \apj, 247, 787

\bibitem[{{Balland} {et~al.}(1998){Balland}, {Silk}, \& {Schaeffer}}]{bal98a}
{Balland}, C., {Silk}, J., \& {Schaeffer}, R. 1998, \apj, 497, 541

\bibitem[{{Bekki}(1999)}]{bek99}
{Bekki}, K. 1999, \apjl, 510, L15

\bibitem[{{Butcher} \& {Oemler}(1984)}]{but84b}
{Butcher}, H. \& {Oemler}, A., J. 1984, \apj, 285, 426

\bibitem[{{Caldwell} \& {Rose}(1997)}]{cal97}
{Caldwell}, N. \& {Rose}, J.~A. 1997, \aj, 113, 492

\bibitem[{{Cesarsky} {et~al.}(1996)}]  {ces96} {Cesarsky}, C.~J., 
{Abergel}, A., {Agnese}, P., {et~al.,} 1996, \aap, 315, L32



\bibitem[{{Condon}(1989)}]{con89}
{Condon}, J.~J. 1989, \apj, 338, 13

\bibitem[{{Condon}(1992)}]{con92}
---. 1992, \araa, 30, 575

\bibitem[{{Condon} {et~al.}(1991){Condon}, {Anderson}, \& {Helou}}]{con91}
{Condon}, J.~J., {Anderson}, M.~L., \& {Helou}, G. 1991, \apj, 376, 95

\bibitem[{{Duc} {et~al.}(2002){Duc}, {Poggianti}, {Fadda}, {Elbaz},
{Flores}, {Chanial}, {Franceschini}, {Moorwood}, \&
{Cesarsky}}]{duc02} {Duc}, P.-A., {Poggianti}, B.~M., {Fadda}, D. and
{Elbaz}, D., {Flores}, H., {Chanial}, P., {Franceschini}, A.,
{Moorwood}, A., \& {Cesarsky}, C.2002, \aap, 382, 60


\bibitem[{{Dwarakanath} \& {Owen}(1999)}]{dwa99}
{Dwarakanath}, K.~S. \& {Owen}, F.~N. 1999, \aj, 118, 625

\bibitem[{{Evrard}(1990)}]{evr90b}
{Evrard}, A.~E. 1989, in Clusters of Galaxies, ed. W.~R. Oegerle, M.~J.
  Fitchett, \& L.~Danly (Cambridge: Cambridge Univ. Press), 287


\bibitem[{{Geller} \& {Beers}(1982)}]{gel82}
{Geller}, M.~J. \& {Beers}, T.~C. 1982, \pasp, 94, 421

\bibitem[{{Girardi} {et~al.}(1995){Girardi}, {Biviano}, {Giuricin},
  {Mardirossian}, \& {Mezzetti}}]{gir95}
{Girardi}, M., {Biviano}, A., {Giuricin}, G., {Mardirossian}, F., \&
  {Mezzetti}, M. 1995, \apj, 438, 527



\bibitem[{{Kennicutt} {et~al.}(1998){Kennicutt}, {Schweizer},
 \& {Barnes}}]{ken98} {Kennicutt}, R.~C., {Schweizer}, F. \& {Barnes},
 J.~E.  1998, in Saas-Fee Advanced Course 26: Galaxies: Interactions
 and Induced Star Formation, 26


\bibitem[{{King}(1962)}]{kin62}
{King}, I. 1962, \aj, 67, 471

\bibitem[{{Lacey} \& {Cole}(1993)}]{lac93}
{Lacey}, C. \& {Cole}, S. 1993, \mnras, 262, 627

\bibitem[{{Lacey} \& {Cole}(1994)}]{lac94}
---. 1994, \mnras, 271, 676

\bibitem[{{Ledlow} \& {Owen}(1995{\natexlab{a}})}]{led95b}
{Ledlow}, M.~J. \& {Owen}, F.~N. 1995{\natexlab{a}}, \aj, 110, 1959

\bibitem[{{Ledlow} \& {Owen}(1995{\natexlab{b}})}]{led95a}
---. 1995{\natexlab{b}}, \aj, 109, 853

\bibitem[{{Ledlow} {et~al.}(1999){Ledlow}, {Owen}, {Dwarakanath}, {Keel}, \&
  {Morrison}}]{led99}
{Ledlow}, M.~J., {Owen}, F.~N., {Dwarakanath}, K.~S., {Keel}, W.~C., \&
  {Morrison}, G.~E. 1999, in ASP Conf. Ser. 176: Observational Cosmology: The
  Development of Galaxy Systems, 83

\bibitem[{{Moore} {et~al.}(1996){Moore}, {Katz}, {Lake}, {Dressler}, \&
  {Oemler}}]{moo96}
{Moore}, B., {Katz}, N., {Lake}, G., {Dressler}, A., \& {Oemler}, A., J. 1996,
  \nat, 379, 613

\bibitem[{{Morrison} {et~al.}(2003{\natexlab{a}}){Morrison}, {Owen}, \&
  {Ledlow}}]{mor00b}
{Morrison}, G.~E., {Owen}, F.~N., \& {Ledlow}, M.~J. 2003{\natexlab{a}}, paper
  III, in preparation

\bibitem[{{Morrison} {et~al.}(2003{\natexlab{b}}){Morrison}, {Owen}, {Ledlow},
  {Hill}, \& {Herter}}]{mor00c}
{Morrison}, G.~E., {Owen}, F.~N., {Ledlow}, M.~J., {Hill}, J., \& {Herter}, T.
  2003{\natexlab{b}}, paper IV, in preparation


\bibitem[{{Morrison} {et~al.}(2002){Morrison}, {Owen}, {Ledlow},
  {Keel}, {Hill}, \& {Voges}}]{mor00} {Morrison}, G.~E., {Owen},
  F.~N., {Ledlow}, M.~J., {Keel}, W.~C., {Hill}, J.~M., \& {Voges},
  W. 2002, paper I, accepted by ApJS

\bibitem[{{Oemler}(1974)}]{oem74}
{Oemler}, A., J. 1974, \apj, 194, 1

\bibitem[{{Oemler} {et~al.}(1997){Oemler}, {Dressler}, \& {Butcher}}]{oem97}
{Oemler}, Augustus, J., {Dressler}, A., \& {Butcher}, H.~R. 1997, \apj, 474,
  561

\bibitem[{{Owen} {et~al.}(1999){Owen}, {Keel}, {Ledlow}, \& {Morrison}}]{owe99}
{Owen}, F.~N., {Keel}, W., {Ledlow}, M.~J., \& {Morrison}, G.~E. 1999, \aj,
  118, 633

\bibitem[{{Sarazin} \& {Quintana}(1985)}]{sar85}
{Sarazin}, C.~L. \& {Quintana}, H. 1985, Galaxy Distribution in Abell Clusters
  (University of Virginia report)


\bibitem[{{Stanford} {et~al.}(1998){Stanford}, {Eisenhardt},\& {Dickinson}}]{sta98}
{Stanford}, S.~A., {Eisenhardt}, P.~R. \& {Dickinson}, M. 1998, \apj, 492, 461

\bibitem[{{Tomita} {et~al.}(1996){Tomita}, {Nakamura}, {Takata}, {Nakanishi},
  {Takeuchi}, {Ohta}, \& {Yamada}}]{tom96}
{Tomita}, A., {Nakamura}, F.~E., {Takata}, T., {Nakanishi}, K., {Takeuchi}, T.,
  {Ohta}, K., \& {Yamada}, T. 1996, \aj, 111, 42

\bibitem[{{Wang} {et~al.}(1997){Wang}, {Ulmer}, \& {Lavery}}]{wan97c}
{Wang}, Q.~D., {Ulmer}, M.~P., \& {Lavery}, R.~J. 1997, \mnras, 288, 702

\bibitem[{{West} \& {Bothun}(1990)}]{wes90}
{West}, M.~J. \& {Bothun}, G.~D. 1990, \apj, 350, 36

\bibitem[{{White}(1976)}]{whi76}
{White}, S. D.~M. 1976, \mnras, 177, 717


\end{thebibliography}

\clearpage

\begin{deluxetable} {ccccc} 
\tablecaption{Radio Selected Galaxy Classes at 1.4\,GHz \label{table_1}}
\startdata\tableline
&\colhead{Class} &\colhead{$\log (L_{\mathrm{min}})$ \whz} & 
\colhead{$\log (L_{\mathrm{max}})$ \whz}&\nl\tableline
&SBRG & 22.3 & 22.75 \nl &LLRG & 22.3& 23 \nl &HLRG & 23& 25 \nl
\enddata
\end{deluxetable}

\clearpage
\centerline{\bf FIGURE CAPTIONS}

\figcaption[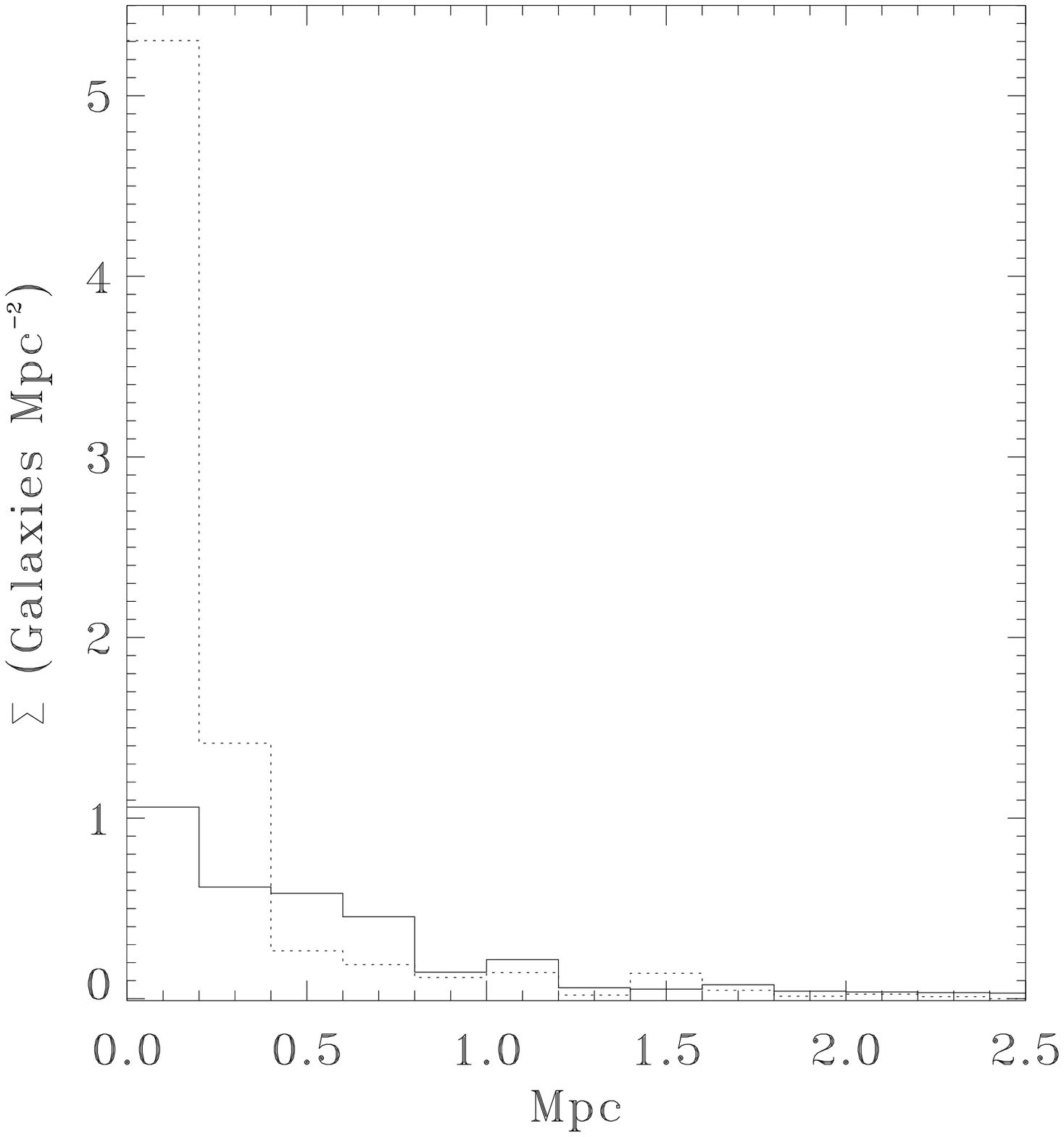]{Projected surface density distribution of radio-selected
galaxies.  Solid line: SBRGs. Dotted line: HLRGs. \label{FIG_1}}

\figcaption[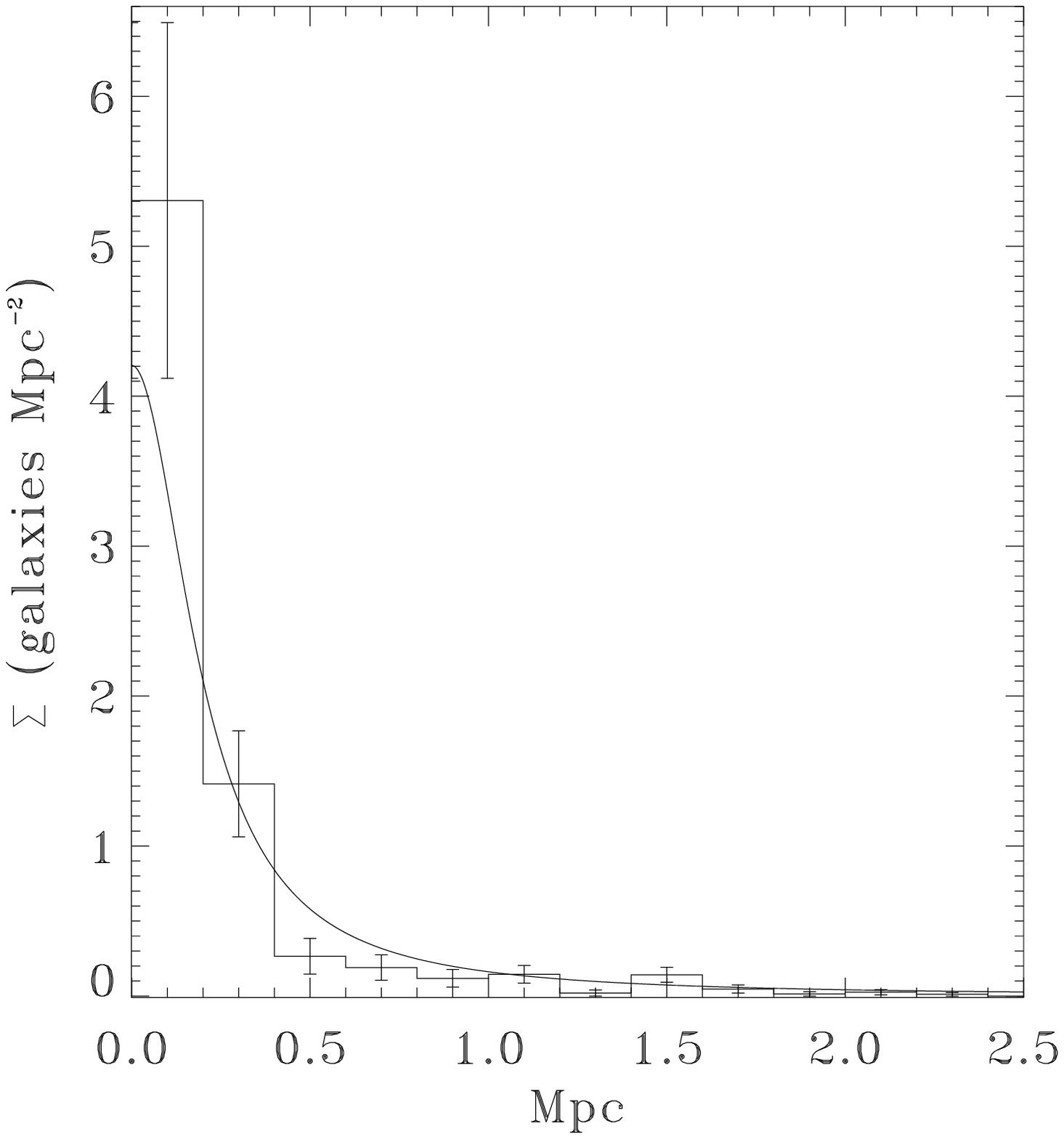]{Projected surface density distribution of high-luminosity
radio-selected galaxies versus the projected radial distance from
cluster center. The curve is a King model, indicating an excess of
galaxies at the center of the cluster. The error bars are Poisson
errors based on the number counts within each bin. \label{FIG_2}}

\figcaption[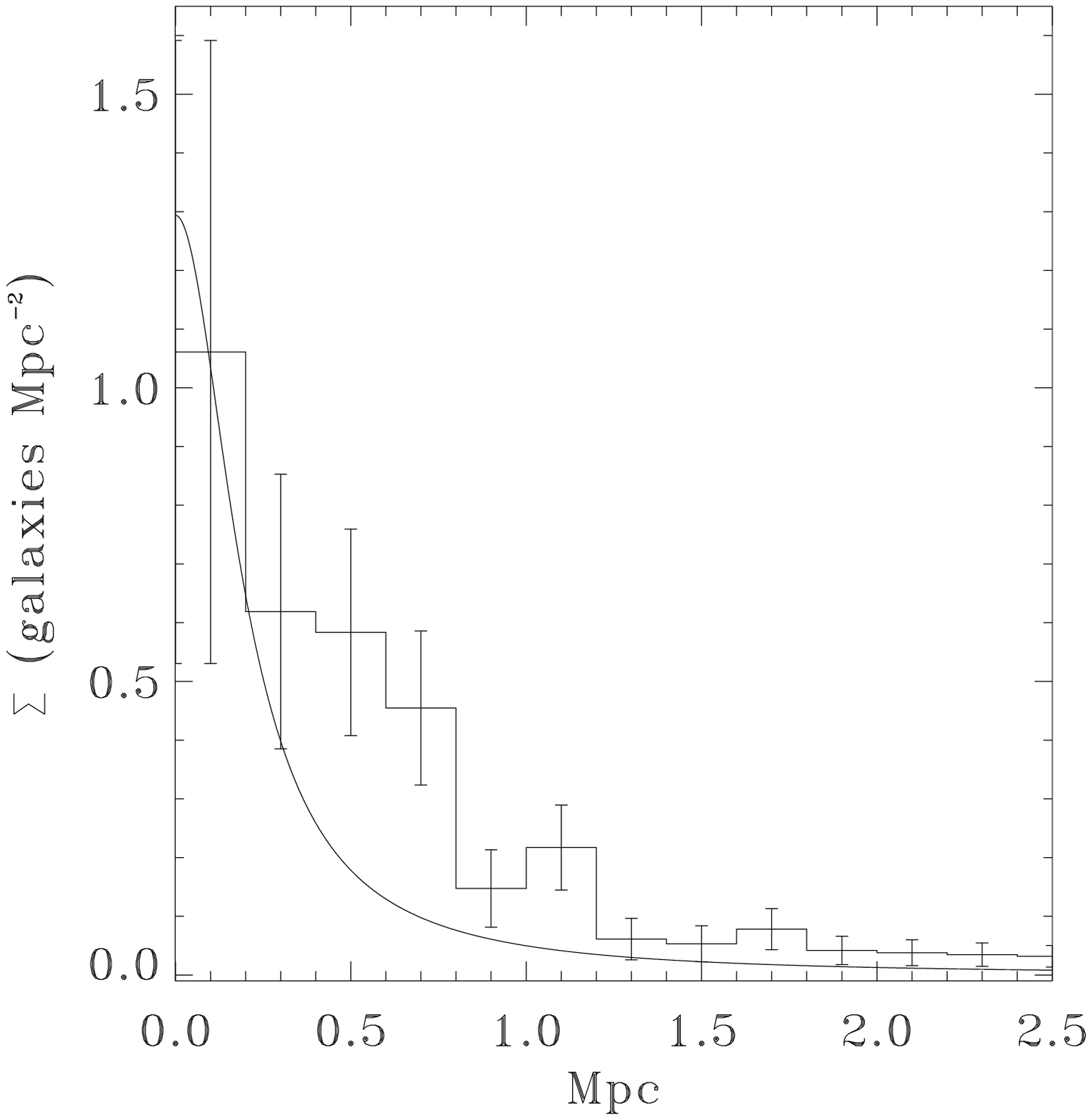]{Same as in Fig. 2 but for the SBRGs. The King model here,
indicates an excess of galaxies beyond $r \gtrsim 0.2$\,Mpc. The
error bars are Poisson errors based on the number counts within each
bin. \label{FIG_3}}

\figcaption[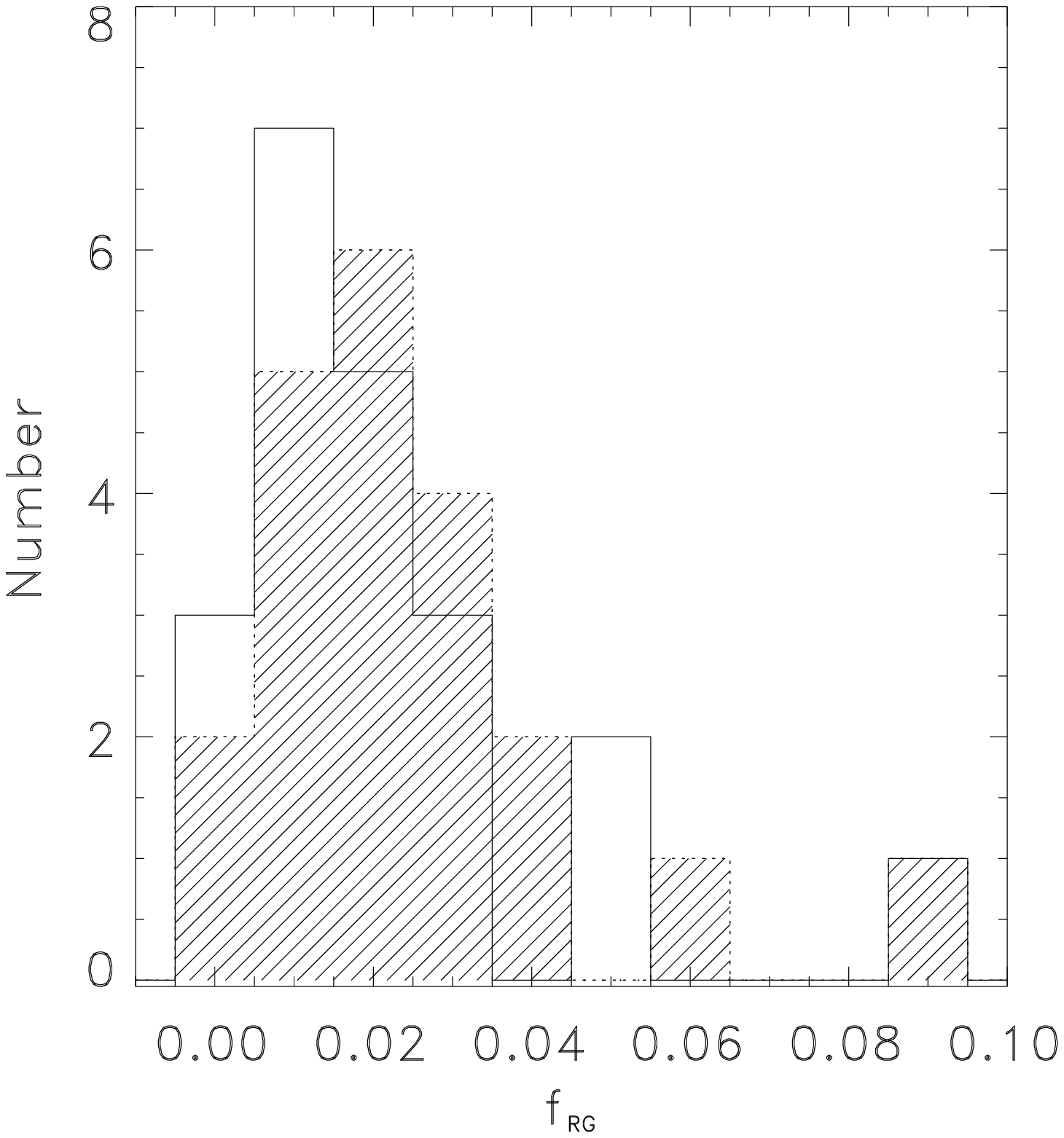]{The SBRGs radio galaxy fractions are
represented by the solid line. A2125 is an outlier at 0.09. The HLRGs
radio galaxy fractions are shown by the hatch patter . A1689 is an
outlier at 0.09 which over lies A2125 data point. \label{FIG_4}}

\figcaption[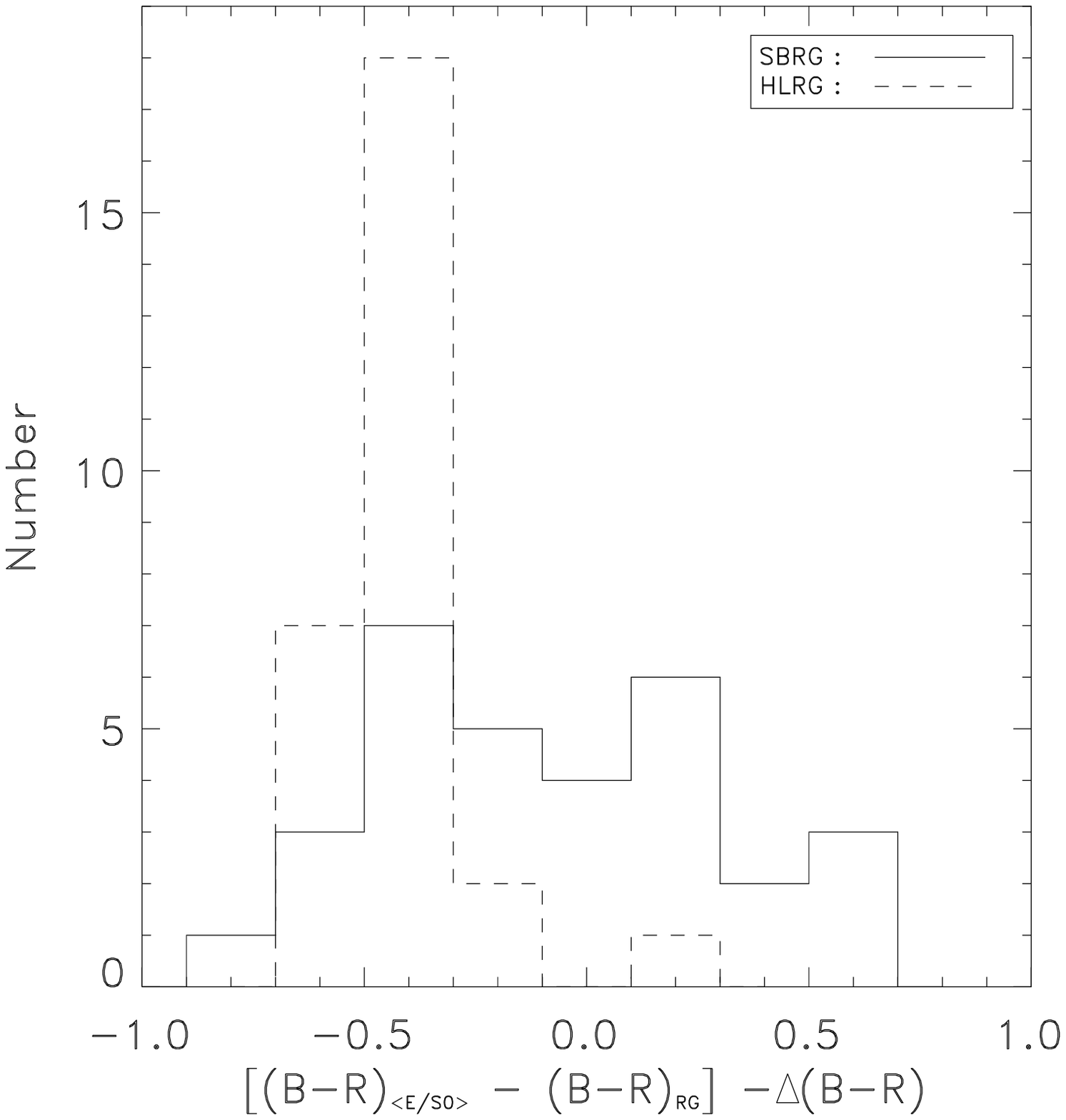]{Histogram of Butcher-Oemler defined blue ($\geq 0$) and red ($<
0$) radio galaxies. Solid line: SBRG. Dashed line: HLRGS.\label{FIG_5}}

\figcaption[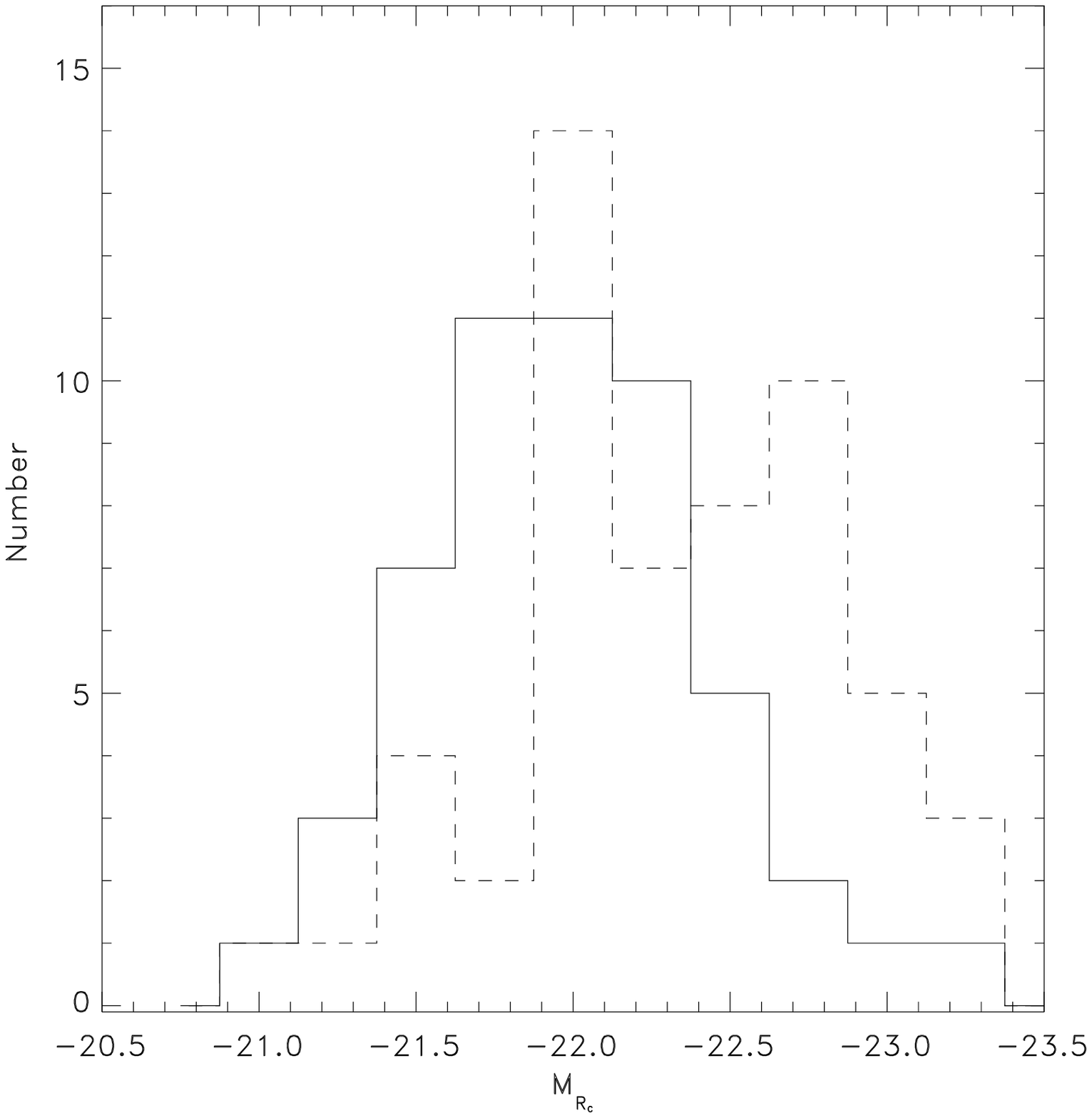]{SBRG (solid line) and HLRG (dashed line) absolute
magnitudes. \label{FIG_6}}

\clearpage

\begin{figure}[p]
\begin{center}
\epsfig{file=Morrison.f1.eps, height=6.0in, width=6.0in}
\end{center}
\end{figure}

\clearpage

\begin{figure}[p]
\begin{center}
\epsfig{file=Morrison.f2.eps, height=6.0in, width=6.0in}
\end{center}
\end{figure}

\clearpage

\begin{figure}[p]
\begin{center}
\epsfig{file=Morrison.f3.eps, height=6.0in, width=6.0in}
\end{center}
\end{figure}

\clearpage

\begin{figure}[p]
\begin{center}
\epsfig{file=Morrison.f4.eps, height=6.0in, width=6.0in}
\end{center}
\end{figure}

\clearpage

\begin{figure}[p]
\begin{center}
\epsfig{file=Morrison.f5.eps, height=6.0in, width=6.0in}
\end{center}
\end{figure}

\clearpage

\begin{figure}[p]
\begin{center}
\epsfig{file=Morrison.f6.eps, height=6.0in, width=6.0in}
\end{center}
\end{figure}

\end{document}